# Athermal resistance to phase interface motion due to precipitates: A phase field study

Mahdi Javanbakht[a*], Valery I. Levitas[b,c]

[a] Department of Mechanical Engineering, Isfahan University of Technology, Isfahan 84156-83111, Iran

[b] Department of Mechanical Engineering, Iowa State University, Ames, IA 50011 USA

[c] Department of Aerospace Engineering, Iowa State University, Ames, IA 50011 USA

**Abstract**

Athermal resistance to the motion of a phase interface due to a precipitate is investigated. The coupled phase field and elasticity equations are solved for the phase transformation (PT). The volumetric misfit strain due to the precipitate is included using the error and rectangular functions. Due to the presence of precipitates, the critical thermal driving forces remarkably differ between the direct and reverse PTs, resulting in a hysteresis behavior. For the precipitate radius small compared to the interface width, the misfit strain does not practically show any effect on the critical thermal driving force.  Also, the critical thermal driving force value nonlinearly increases vs. the precipitate concentration for both the direct and reverse PTs. Change in the precipitate surface energy significantly changes the PT morphology and the critical thermal driving forces. The critical thermal driving force shows dependence on the misfit strain for large precipitate sizes compared to the interface width. For both the constant surface energy (CSE) and variable surface energy (VSE) boundary conditions (BCs) at the precipitate surface, the critical thermal driving force linearly increases vs. the misfit strain coefficient for the direct PT while it is almost independent of it for the reverse PT. For larger precipitates, the critical thermal driving force nonlinearly increases vs. the precipitate concentration for the direct PT.  For the reverse PT, however, its value for the CSE BCs linearly increases vs. the precipitate concentration while it is almost independent of the precipitate concentration for the VSE BCs. Also, for any concentration, the VSE BCs result in higher thermal critical driving forces, a smaller hysteresis range, and a larger transformation rate. The obtained critical microstructure and thermal driving forces are validated using the thermodynamic phase equilibrium condition for stationary interfaces. The obtained results help for a better understanding of athermal friction mechanism for interfaces and similar defect effects on various PTs at the nanoscale.

# 1. Introduction

The interaction of PTs with structural defects plays a crucial role in determining material properties. Precipitates can significantly affect the thermodynamics, kinetics, and morphology of martensitic PTs. The effect of precipitates on martensitic PT in a b-CuAlBe SMA was experimentally studied in [1] which showed a significant change in morphology and thermodynamic conditions. The effect of $Ni_4Ti_3$ precipitates on martensitic PTs and consequently, the shape memory effect in TiNi was studied in [2,3]. The effect of defects on austenite-martensite interfaces was investigated in CuAlNi which revealed the significant influence on thermoelastic first order PT equilibrium [4]. The significant effect of $Ni_4Ti_3$ nanoprecipitates on the stress- and temperature-induced PTs, superelastic hysteresis loop, transformation temperatures and other characteristics of NiTi was found using MD simulations [5]. Martensitic transformation induced by the precipitate was studied using phase field approach [6]. An internal friction model was suggested based on the theory of phase nucleation and growth which shows the significant change in the thermoelastic martensitic transformation due to interstitials [7]. At nanoscale, athermal friction can be caused by the Peierls barrier or the interaction of interface with long-range stress fields of defects [8-10] which can change the nanostructure evolution, kinetics and thermodynamics and is responsible for temperature hysteresis. The athermal threshold in the PFA was included by introducing oscillating stress fields due to the Peierls barrier and defects or a jump in chemical energy [10,11]. The athermal hysteresis was modeled due to the dislocations generated along the austenite-martensite interface [12]. Such hysteresis behavior is even more crucial for PTs from the low-pressure phases (LPP) to high-pressure phases (HPPs) due to the large range of transformation pressures. HPPs (e.g., diamond and cubic BN) may have desired physical properties. One of the goals in the synthesis of HPPs is to reduce PT pressure for direct PT and to suppress the reverse PT, so that HPPs can be used at normal pressure in engineering applications. That is why controlling athermal interface friction by precipitates can play a key role on the LPP-HPP transformation characteristics in various materials. Many numerical and experimental studies have been done to discover new HPPs and their PT mechanisms [13-15]. Recent reviews that includes analysis of hysteretic behavior for HP PTs are presented in [16,17]. Despite the extensive research on HPPs, the athermal friction to the interface motion and the hysteresis in HPPs due to precipitates has not been studied yet, and this is the main focus of the current work.

It should be noted that structural defects can create a significant heterogeneity and lead to high stress concentrations and consequently, contribute to the thermodynamic driving forces for PT. This can promote direct or reverse LPP-HPP transformations at low mechanical loadings and even without them [18-22]. Therefore, even without mechanical loading, PTs can occur due to change in temperature (thermally-induced PT) and the internal stresses due to stress concentrators caused by structural defects such as precipitates and inelastic transformation strain mismatch at the interfaces. Since our focus is to find the athermal interface resistance, we will not apply external loading, but thermal hysteresis can be recalculated into pressure hysteresis using known equations. There exist various analytical, numerical, and experimental works on the effect of particles and precipitates on grain boundary (GB) motion. 3D FEM simulations for GB motion through spherical incoherent particles were presented [23]. An experimentally proven model for predicting austenite grain growth including the pinning and solute-drag effects of TiN precipitates and assuming a constant width of an austenite grain boundary was proposed in [24]. The effect of the interfacial energy of grain/particle was studied on grain growth kinetics using a cellular automata model [25]. The interaction between Ta clusters and GBs was studied in nanocrystalline Cu-Ta alloys which revealed the key role of temperature and composition [26]. A random walk-based model for GB motion through particles was introduced [27] which attributes GB fluctuations to the boundary mobility and drag effect. Molecular dynamics simulations of pinning of a Cu GB by an Ag particle have been presented in [28]. Interactions between GBs and Pt particles in Pt-implanted high-purity polycrystalline $Al_2O_3$ were experimentally studied revealing various transitional morphologies [29]. The transmission electron microscopy and tensile tests showed that both the solute drag effect and the precipitate Zener effect are responsible for the retarded recrystallization in Mo-modified Zr-Nb alloys, which also significantly affected the grain size, yield strength, and ductility [30]. The kinetics of austenite grain growth was predicted in agreement with experiments using both the precipitation and grain growth models [31]. The effect of particles and precipitates on the interfaces is also investigated in few works. A relationship in terms of nanoparticles size and distribution for the Zener pressure was described in [32]. A significant Zener effect was found during annealing in Al–Mg–Si alloy due to the high density of the $L1_2$ dispersoids [33]. The interaction between interfaces and precipitated carbides was experimentally studied. which attributed the shifts in the critical temperatures to the pinning force [34].

Within the continuum modeling of PTs, phase field approach (PFA) has been effectively used to predict structural evolution at various size scales [35] and particularly at the nanoscale for crack [36], nanovoids [37,38], grains [39], martensitic PTs [40-43] and dislocations [44-47]. The PFA deals with an energy function in terms of order parameters and their gradients. The microstructure evolution is described by corresponding thermodynamically consistent kinetic equation.

Besides the various works above, the PFA has been recently used for modeling the effect of particles and precipitates on GBs and interfaces. 3D PF simulations of moving GBs through cylindrical particles in composites were presented and the effect of relative orientation and aspect ratio of particles on the kinetics was studied [48]. 3D PF simulations for the grain growth were performed which revealed the effect of particle-matrix coherency on GB pinning [49]. A PFA was proposed to study the effect of coherent precipitate on the Zener pinning of GBs which includes the misfit strain and the elastic heterogeneity and anisotropy [50]. The PF simulations were performed for 2D and 3D polycrystalline materials, and the pinning effect of incoherent particles on GBs was studied [51]. Within the PFA, an order parameter ($\eta$) is defined which varies from 0 for the parent phase (which we will call LPP, while it is equally applied to temperature-induced PT) to 1 for HPP. The PT conditions for the model that we will used were derived first in [52-54]. Surface-induced PTs and interface stresses were presented in [55-58]. The interaction of PTs with dislocations [18-20], voids [21], and cracks [59] revealed a significant change in the PT mechanism and conditions. The FPA for martensitic PTs was reviewed in [60].

In the current paper, we present the first detailed PFA study of the athermal friction to the interface motion caused by a precipitate, which includes two different profiles of a mismatch strain within an interface, unchanged and changed precipitate surface energy during the PT, and analysis their effect, as well as precipitate size, phase interface width, and volume fraction of the precipitates effects. The paper is organized as follows. In Section 2, the phase field model as well as the precipitate model are presented. The numerical procedure is described in Section 3. The results including critical thermal driving forces vs. precipitate concentration and misfit strain for direct and reverse PTs and corresponding nanostructures, hysteresis region, nanostructure evolution during direct and reverse PTs, size effect, interface width effect and the effect of precipitate variable surface energy on the critical driving forces are presented with their discussion in Section 4. Concluding remarks are summarized in Section 5.

## 2. System of equations

### 2.1. Phase field model for LPP-HPP transformation [18,20]

The transformation between the LPP and the HPP is described using the Ginzburg-Landau (GL) equation for the evolution of the order parameter $\eta$ as

$$\frac{1}{\lambda}\frac{\partial \eta}{\partial t} = -\frac{\partial \psi_P}{\partial \eta}\Big|_\varepsilon + \beta \nabla^2 \eta. \tag{1}$$

The Helmholtz free energy per unit volume $\psi_P$ is defined as

$$\psi_P = \frac{1}{2}\boldsymbol{\varepsilon}_e : \boldsymbol{C} : \boldsymbol{\varepsilon}_e + A_0(\theta - \theta_c)\eta^2(1-\eta)^2 + z(\theta - \theta_e)\eta^3(4-3\eta) + \frac{\beta}{2}|\nabla \eta|^2. \tag{2}$$

Here, $\theta$ is the temperature, $\theta_e$ is the phase equilibrium temperature at zero stresses, $\theta_c$ is the critical temperature for the loss of stability of the stress-free LPP, $\boldsymbol{\varepsilon}_e$ is the elastic strain tensor, and $\boldsymbol{C}$ is the tensor of elastic moduli. The material parameters $\lambda$, $\beta$, $z$, and $A_0$ are defined in Table. 1. For these parameters, the interface width $\delta_0 = 5.54\sqrt{\beta/(2A_0(\theta_e - \theta_c))} = 1.43$ nm and energy $\gamma = \sqrt{\beta A_0(\theta_e - \theta_c)/18} = 0.36\ J/m^2$. The transformation strain tensor $\boldsymbol{\varepsilon}_t$ varies from zero for LPP to $\boldsymbol{\varepsilon}_{tr}$ for HPP as

$$\boldsymbol{\varepsilon}_t = \boldsymbol{\varepsilon}_{tr}[a\eta^2 + (4-2a)\eta^3 + (a-3)\eta^4], \tag{3}$$

where $a$ is a material parameter which characterizes the PT equilibrium and instability pressures [52] and will be determined below. For the total strain tensor $\boldsymbol{\varepsilon}$ we accept the additive decomposition of elastic ($\boldsymbol{\varepsilon}_e$), transformation ($\boldsymbol{\varepsilon}_t$), and precipitate misfit ($\boldsymbol{\varepsilon}_{pr}$) strain tensors as

$$\boldsymbol{\varepsilon} = \boldsymbol{\varepsilon}_e + \boldsymbol{\varepsilon}_t + \boldsymbol{\varepsilon}_{pr}. \tag{4}$$

For convenience of describing the material in terms of characteristic pressures instead of temperatures, we introduce the phase equilibrium pressure, $p_e$, and the lattice instability pressures for the direct PT, $p_{in}$, and reverse PT, $p_{in}^r$, as [18,20,52,53]

$$p_e = \frac{z(\theta - \theta_e)}{\varepsilon_0}; \tag{5}$$

$$p_{in} = \frac{A_0(\theta - \theta_c)}{a\varepsilon_0};$$

$$p_{in}^r = \frac{6z(\theta - \theta_e) - A_0(\theta - \theta_c)}{(6-a)\varepsilon_0}.$$

We accept $p_e = 10$, $p_{in} = 20$, and $p_{in}^r = -10$ at $\theta = 300K$ [38]. For the plain-strain formulation, the volumetric transformation strain $\varepsilon_0 = (\varepsilon_{trx} + \varepsilon_{try}) = -0.1$, see Table 1. Substituting these values in Eq. (5) gives $a = 4$, $A_0 = 20.6 MPa$, and $z = 5.05 MPaK^{-1}$. The elasticity equations which will be coupled to the GL equation are

$$\nabla \cdot \boldsymbol{\sigma} = \boldsymbol{0};$$

$$\boldsymbol{\varepsilon} = \boldsymbol{\varepsilon}_e + \boldsymbol{\varepsilon}_t + \boldsymbol{\varepsilon}_{pr} = \frac{1}{2}[\nabla \boldsymbol{u} + (\nabla \boldsymbol{u})^T]; \tag{6}$$

$$\boldsymbol{\sigma} = \boldsymbol{C}:(\boldsymbol{\varepsilon} - \boldsymbol{\varepsilon}_t - \boldsymbol{\varepsilon}_{pr}),$$

where $\boldsymbol{\sigma}$ is the elastic stress tensor and $\boldsymbol{u}$ is the displacement field. Substituting Eqs. (3) and (6) with $a = 4$ into Eq. 1 gives the GL equation as

$$\frac{1}{\lambda}\frac{\partial \eta}{\partial t} = 4\eta(\eta - 1)(\eta - 2)\boldsymbol{\sigma}:\boldsymbol{\varepsilon}_{tr} \\ - [2A_0(\theta - \theta_c)\eta(\eta - 1)(2\eta - 1) + 12z(\theta - \theta_e)\eta^2(1 - \eta)] + \beta \nabla^2 \eta. \tag{7}$$

The insulated boundary condition for the PT problem is $\beta \nabla \eta \cdot \boldsymbol{n} = 0$, where $\boldsymbol{n}$ is the normal to the boundary. It means that the surface energy of the external boundaries does not change during the PT.

### 2.2. Precipitate model

The precipitate is modeled as a non-evolving circular region inside which no PT occurs and it includes a misfit strain due to the compositional heterogeneity between the matrix and precipitate [50,61]. The corresponding misfit strain tensor with respect to LPP is modeled as a position dependent volumetric strain $\boldsymbol{\varepsilon}_{pr} = \varepsilon_p(\boldsymbol{x})\boldsymbol{I}$ [50], where $\boldsymbol{x}$ is the position vector and $\boldsymbol{I}$ is the unit tensor. The position dependence or the distribution of $\varepsilon_p(\boldsymbol{x})$ is considered using two different models: (a) a jump function such that $\varepsilon_p(\boldsymbol{x}) = \varepsilon_v$ inside the precipitate region and it is zero in the rest of sample, where $\varepsilon_v$ is the misfit strain coefficient, and (b) the error function $\varepsilon_p(\boldsymbol{x}) = \varepsilon_v/(\pi l^2)\exp(-|\boldsymbol{x}' - \boldsymbol{x}|^2/l^2)$, where $l$ is the radius of the circular region with the center $\boldsymbol{x}'$, within which $\varepsilon_p(\boldsymbol{x})$ is nonzero [62] and smoothly goes to zero outside of it. Our results show practically no difference between these two models due to the nanoscale size of the precipitate. The precipitate concentration, $c$, is defined as the ratio of the area of the precipitate to that of the entire sample.

The misfit strain is assumed to be independent of the order parameter. Here, a range of compositional heterogeneity due to different types of precipitate is characterized by the range $0 \leq \varepsilon_v \leq 0.1$ for the misfit strain. In practice, due to its additivity with volumetric part of the transformation strain, this means that the misfit constant with respect to the HPP is $\varepsilon_v + 0.5\varepsilon_0$. Since misfit strain is considered to be tensile, it produces internal compressive mean stress (pressure) in the precipitate and tensile mean stress in the matrix. These stresses suppress direct and promote reverse PTs. Note that the study of the effect of misfit strain on grain growth and GB motion was presented in [50,61], where it was considered independent of GBs.

The boundary between the precipitate and matrix is coherent, i.e., displacements are continuous across the interface. The surface energy of the precipitate during the PT can vary or be constant. The variable surface energy boundary conditions (VSE BCs) between the precipitate and the matrix [6,55,56] are defined as

$$\beta \nabla \eta \cdot \boldsymbol{n} = -\frac{dq}{d\eta}, \quad q = \gamma_{LPP} + (\gamma_{HPP} - \gamma_{LPP})[a\eta^2 + (4 - 2a)\eta^3 + (a - 3)\eta^4], \tag{8}$$

where $q$ is the surface energy during the PT, and $\gamma_{LPP}$ and $\gamma_{HPP}$ are the surface energies between the precipitate and the LPP and HPP, respectively. For the constant surface energy (CSE) BCs, $\gamma_{HPP} = \gamma_{LPP}$ and $\beta \nabla \eta \cdot \boldsymbol{n} = 0$. Since $\nabla \eta$ describes the normal to the LPP-HPP interface, then this interface is orthogonal to the precipitate surface. We accept for VSE BCs that $\gamma_{LPP} = 1 \ J/m^3$ and $\gamma_{HPP} = 0.6 \ J/m^3$. The reduction in surface energy promotes HPP and suppresses LPP.

## 3. Numerical procedure

The FEM COMSOL code is used to solve the coupled GL and elasticity equations in 2D, which are implemented in PDE/Heat Transfer in Solids application and the Structural mechanics/Plane Strain application, respectively. Triangular Lagrange elements with the mesh size of 0.5nm are used to reach mesh-independent solutions. The Segregated solver with the time step of 0.01ps has been utilized. Stress, size, and time are normalized by 1 GPa, 1 nm, and 1 ps, respectively. The material parameters for the PT simulations [18,20] are given in Table 1. The numerical solutions well resolve the analytical solutions for the planar austenite-martensite interface energy and width [18,19,22].

Table 1. The material parameters used in the PT simulations.

| Parameter | Value | Definition |
|---|---|---|
| $A_0$ | 20.6 MPaK$^{-1}$ | The magnitude of the double well barrier between LPP-HPP |
| $z$ | $-5.05$ MPaK$^{-1}$ | The jump in specific entropy |
| $\beta$ | $5.18 \times 10^{-10}$ N | LPP-HPP gradient energy coefficient |
| $\lambda$ | 2600 (Pa.s)$^{-1}$ | Kinetic coefficient |
| $\theta_e$ | 100 K | Phase equilibrium temperature at zero stresses |
| $\theta_c$ | $-90$ K | Critical temperature for the loss of stability of the stress-free LPP |
| $E$ | 177.023 GPa | Young's modulus |
| $\nu$ | 0.238 | Poisson's ratio |
| $\varepsilon_{tr}$ | $\begin{vmatrix} -.05 & 0.1 \\ 0.1 & -.05 \end{vmatrix}$ | Transformation strain tensor |

## 4. Results

The athermal resistance to the LPP-HPP interface motion due to precipitates is investigated. A circular precipitate region with the radius of R is located at the center of a square sample with the size L. The lower left corner is fixed in both x and y directions and the upper left corner is fixed only in the x direction. Initially, to avoid nucleation problem, a small part of the left side of the sample is considered a HPP and the rest of it is considered a LPP. At low temperatures, the initial sharp interface between the HPP and LPP broadens to a diffuse interface which propagates to the right (direct PT), while at high temperatures a reverse PT occurs, and the interface moves back to the left. In the absence of the precipitate, slightly below the phase equilibrium temperature $\theta_e$, the interface propagates to the right and slightly above $\theta_e$ it moves back to the left. Thus, there is no athermal resistance for direct and reverse PTs and there exists no athermal hysteresis. Due to the presence of a precipitate, the motion of an LPP-HHP interface experiences athermal resistance. Thus, a larger dimensionless thermal driving force $\bar{\theta} = (\theta_e - \theta)/\theta_e$ (or equivalent dimensionless mechanical driving force $\bar{p} = (p - p_e)/p_e$ for the loading with pressure) is required for the interface motion to continue to the right during the direct PT. Conversely, a smaller negative thermal driving force is required to allow the interface pass through the precipitate to the left during the reverse PT. Investigating such hysteresis is the main focus of the current study.

The critical thermal driving force for the direct PT, $\bar{\theta}_c^d$ (blue lines), and reverse PTs, $\bar{\theta}_c^r$ (red lines), are plotted vs. the misfit strain coefficient in Fig. 1a for the misfit strain distribution using the error function and in Fig. 1b for the constant misfit strain. They are defined as the temperatures for which stationary solutions with two-phase regions cease to exist, and solution evolve to the complete HPP for direct PT or complete LPP for the reverse PT. Between $\bar{\theta}_c^d$ and $\bar{\theta}_c^r$, two-phase equilibrium is arrested, which exhibit itself as an athermal resistance to the interface motion. In these simulations, L=20 and R=0.5 (i.e., $c = 0.002$), which is much smaller than the interface width of $\delta_0 = 1.43$. An important finding here is that the misfit strain does not practically show any effect on the critical thermal driving forces for the direct and reverse PTs provided that the precipitate radius size is small compared to the interface width. The difference between the critical thermal driving forces of direct and reverse PTs, $H = \bar{\theta}_c^d - \bar{\theta}_c^r$, defines the athermal hysteresis. Since the counterpart of the phase equilibrium temperature for the system with precipitate cannot be determined, we assume that it is in the middle between $\bar{\theta}_c^d$ and $\bar{\theta}_c^r$, and the athermal resistance to the interface motion is the same in both direction and equal to *0.5H*.

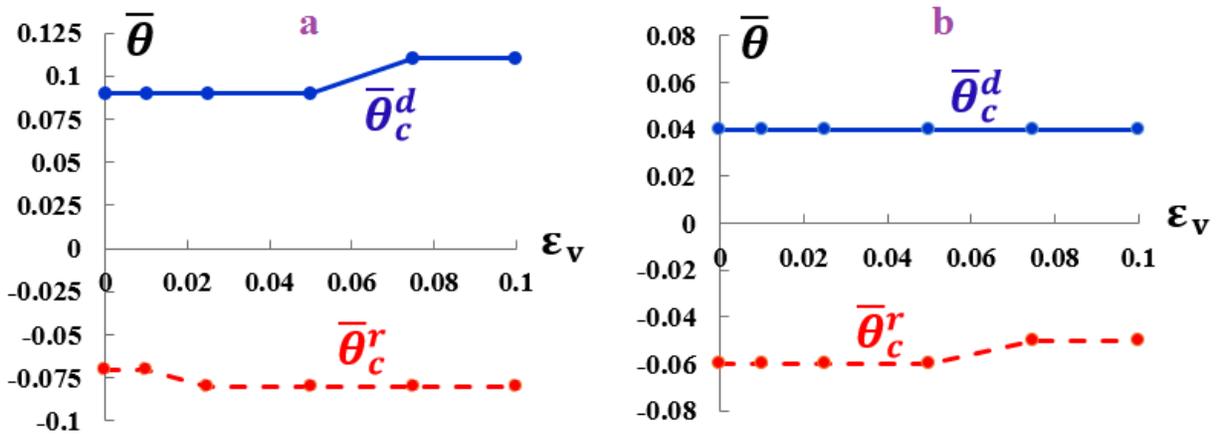

Fig. 1. The critical thermal driving forces for direct (blue lines) and reverse PTs (red lines) vs. the misfit strain coefficient for (a) the misfit strain distribution using the error function and (b) for the constant misfit strain; L=20 and R=0.5, i.e., $c = 0.002$. There is clear effect of the shape of $\varepsilon_v$.

By varying the sample size L in the range of 5 to 50 at constant R=2, the critical thermal driving forces and the hysteresis region are obtained vs. the precipitate concentration (Fig. 2). Obviously, point (0,0) is added for the sample without precipitate. The critical thermal driving forces nonlinearly increases vs. the precipitate concentration for both the direct and reverse PTs, especially for low precipitate concentrations. The critical thermal driving forces for the direct and reverse PTs are approximately symmetric with respect to $\bar{\theta} = 0$ up to $c = 0.015$ and asymmetric above it. The asymmetry is due to the small sample as well as the compressive diagonal transformation strains of the HPP ($\varepsilon_{tr} = \begin{bmatrix} -.05 & 0.1 \\ 0.1 & -.05 \end{bmatrix}$) which produces tensile mean stress in the matrix and promotes reverse PT and suppresses the direct PT. Indeed, for zero diagonal transformation strains, i.e., $\varepsilon_{tr}^* = \begin{bmatrix} 0 & 0.1 \\ 0.1 & 0 \end{bmatrix}$, the critical thermal driving forces $\bar{\theta}_c^d$ and $\bar{\theta}_c^r$ are almost symmetric with respect to $\bar{\theta} = 0$. Note that plots in Fig. 2 are almost the same for any $\varepsilon_v$ in the range $0 \leq \varepsilon_v \leq 0.1$, as expected from Fig. 1, while the critical thermal driving force depends on the misfit strain for larger precipitate sizes, which will be discussed later.

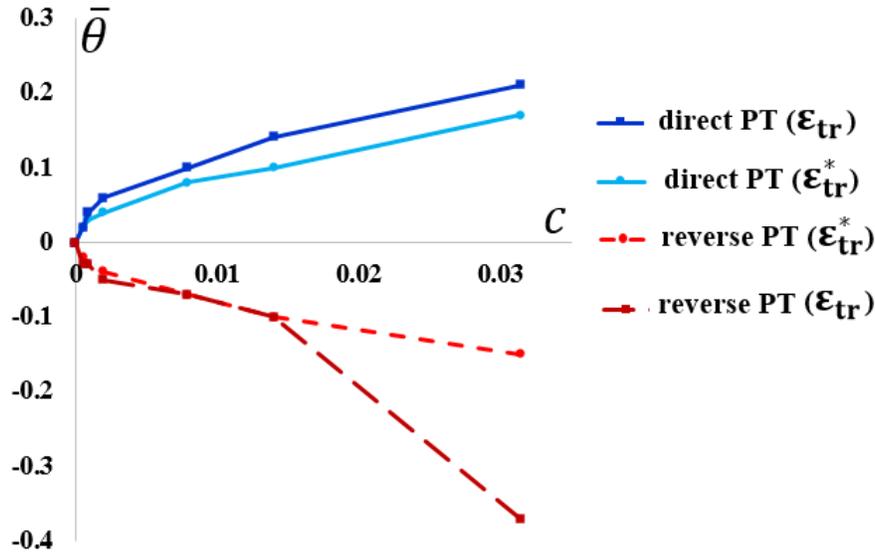

Fig. 2. The critical thermal driving forces vs. the precipitate concentration for the direct and reverse PTs for two different HPP transformation strain tensors, CSE BCs for R=2 and $\varepsilon_v = 0.05$.

Now, let us consider a precipitate with R=2 and $\varepsilon_v = 0.05$. Fig. 3 presents the evolution of the HPP phase for the critical thermal driving force for the direct PT $\bar{\theta}_{th}^d = 0.21$ (a), for a slightly larger thermal driving force $\bar{\theta} = 0.22$ (b) and during the reverse PT for a slightly smaller thermal driving force $\bar{\theta} = -0.12$ than the critical thermal driving force $\bar{\theta}_{th}^r = -0.11$, when the CSE BCs are applied on the precipitate surface. For $\bar{\theta}_{th}^d = 0.21$, the LPP-HPP interface propagates to the right until it reaches the precipitate and stuck, and only slightly rotates at the upper side until it reaches the stationary solution at $t = 40$ (Fig. 3a). Interface below the precipitate is delayed in comparison with the interface above the precipitate due to change in geometry caused by the transformation shear. After increasing the thermal driving force to $\bar{\theta} = 0.22$, the interface passes through the precipitate region and completes the PT to the HPP in the entire sample at $t = 100$ (Fig. 3b). The main event that determines unlimited interface motion is the loss of the stability of the stationary interface near the lower portion of the precipitate. This will be discussed later using the transformation work distribution. Due to the small sample size and the boundary effects, the interface significantly rotates around the precipitate before it leaves its surface. For larger sample sizes, the interface passes the precipitate region without changing its orientation.

For the reverse PT, to avoid nucleation problem like the one for the direct PT, we start simulations before complete direct PT occurs. Here, the solution of Fig. 3b at $t = 90$ is chosen as the initial condition for the reverse PT problem as shown in Fig. 3c. Again, the main event that determines unlimited interface motion is the loss of the stability of the stationary interface near the lower portion of the precipitate, which is clearly delayed in Fig. 3c at t=110 and 120 in comparison with that in the upper portion of the precipitate.

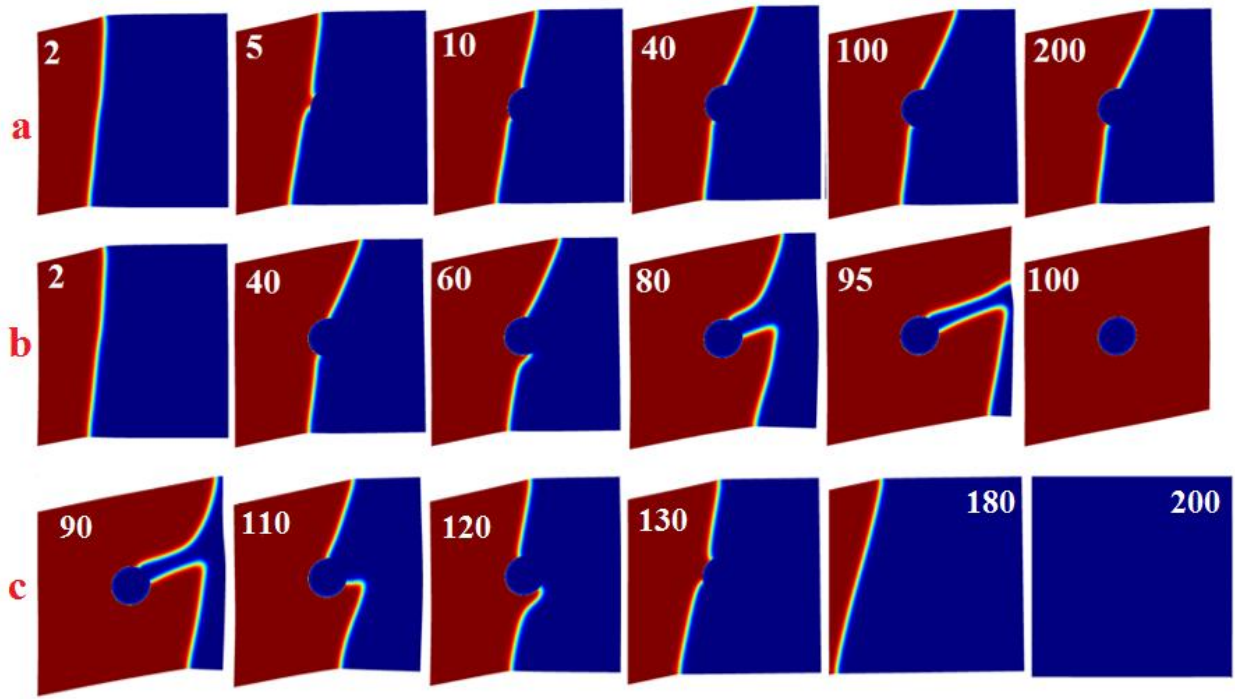

Fig. 3. The evolution of the HPP phase for the critical thermal driving force $\bar{\theta}_{th}^d = 0.21$ up to stationary solution (a), for a slightly larger thermal driving force $\bar{\theta} = 0.22$ (b), and for the thermal driving force $\bar{\theta} = -0.12$ slightly smaller than the critical thermal driving force $\bar{\theta}_{th}^r = -0.11$ for the reverse PT (c), all for CSE BCs at the precipitate surface. L= 20, R=2, and $\varepsilon_v = 0.05$.

As it will be shown, variation in the precipitate surface energy changes the stress distribution and consequently transformation work and can significantly change the PT morphology and the critical thermal driving forces for both the direct and reverse PTs. Thus, the VSE is one of the key parameters in determining the critical thermal driving forces. Fig. 4 presents the evolution of the HPP phase (a) for direct PT for $\bar{\theta} = \bar{\theta}_c^d = 0.28$ up to the stationary solution and (b) for a slightly larger thermal driving force $\bar{\theta} = 0.29$, as well as (c) during the reverse PT for a slightly smaller thermal driving force $\bar{\theta} = -0.02$ than the critical thermal driving force $\bar{\theta}_c^r = -0.01$ for the reverse PT, when the VSE BCs are applied on the precipitate surface. For VSE BCs, the surface energy of the boundary between the precipitate and the matrix varies from $1\ J/m^3$ (for the LPP) to $0.6\ J/m^3$ (for the HPP) in a very thin region with the width of $1\ nm$, while for CSE there is a practically sharp interface between precipitate and matrix for any phase. This significantly changes the stress distribution. For $\bar{\theta} = \bar{\theta}_{th} = 0.28$, after the LPP-HPP interface is arrested, its middle part coincides with the precipitate surface and its upper part rotates by $60^0$ until it reaches the stationary solution at $t = 200$ (Fig. 4a).

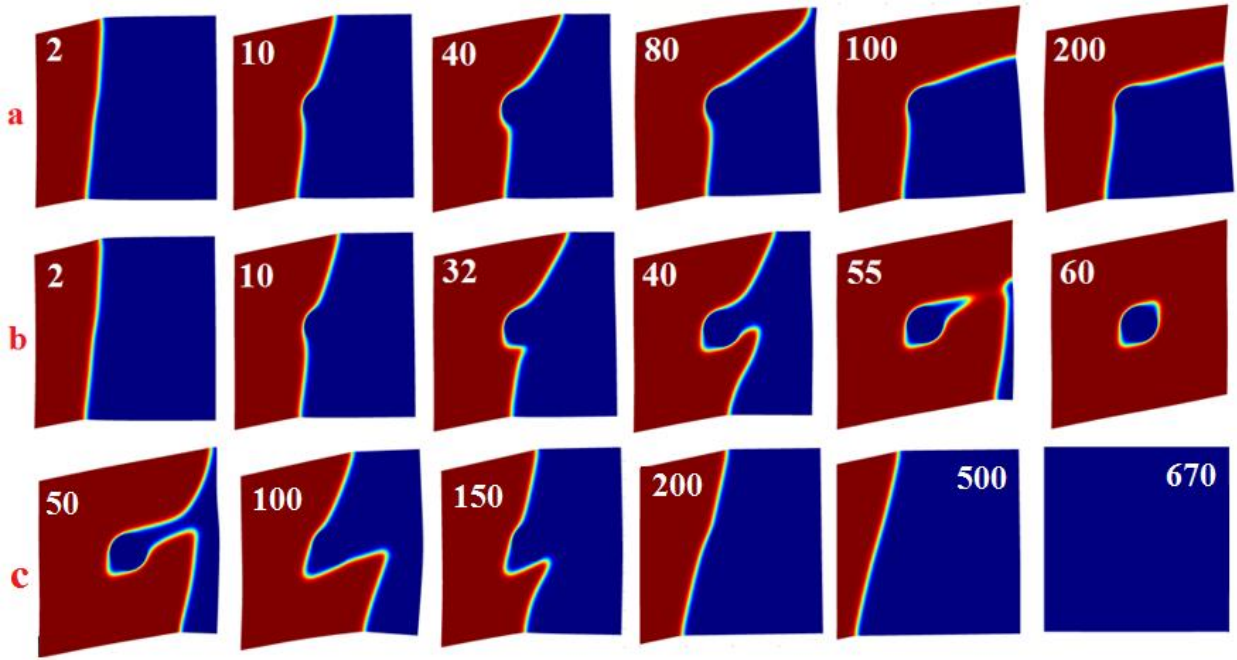

Fig. 4. The evolution of the HPP phase for the critical thermal driving force $\bar{\theta} = \bar{\theta}_{th}^d = 0.28$ up to stationary solution (a), for a slightly larger thermal driving force $\bar{\theta} = 0.29$ (b), and for the thermal driving force $\bar{\theta} = -0.02$, slightly smaller than the critical thermal driving force $\bar{\theta}_{th}^r = -0.01$ for the reverse PT (c) for VSE BCs at the precipitate surface. L= 20, R=2, and $\varepsilon_v = 0.05$.

Slightly increasing the thermal driving force to $\bar{\theta} = 0.29$ results in a very different morphology so that the upper and lower parts of the interface continue propagation until they coalesce and move away from the precipitate (Fig. 4b). Similar to the discussion for Fig. 3, a, non-complete transformed solution in Fig. 4b for $t = 50$ is chosen as the initial condition for the reverse PT. In contrast to the CSE BCs, the reverse PT reveals a different evolution compared to the direct PT, especially when passing the precipitate region (Fig. 4c).

The critical thermal driving forces vs. the misfit strain coefficient are plotted for both the direct and reverse PTs for R=2 (i.e., for the precipitate size larger than the interface width) in Fig. 5. For both the CSE and VSE BCs, the critical thermal driving force linearly increases vs. the misfit strain coefficient for the direct PT while it is almost independent of the misfit strain coefficient for the reverse PT. The VSE BCs also result in higher critical thermal driving forces for both the direct and reverse PTs and a smaller hysteresis compared to those of the CSE BCs. For example, for $\varepsilon_v = 0.08$, $\bar{\theta}_c^d = 0.31$ and $\bar{\theta}_c^r = 0$ for the VSE which gives the hysteresis $H_{VSE} = 0.31$ while for the CSE BCs, $\bar{\theta}_c^d = 0.26$ and $\bar{\theta}_c^r = -0.1$ with the $H_{CSE} = 0.36$. In comparison with Fig. 1, increase in R increases athermal hysteresis and suppress direct PT more than the reverse PT, due to stronger effect of volumetric transformation strain combined with misfit strain.

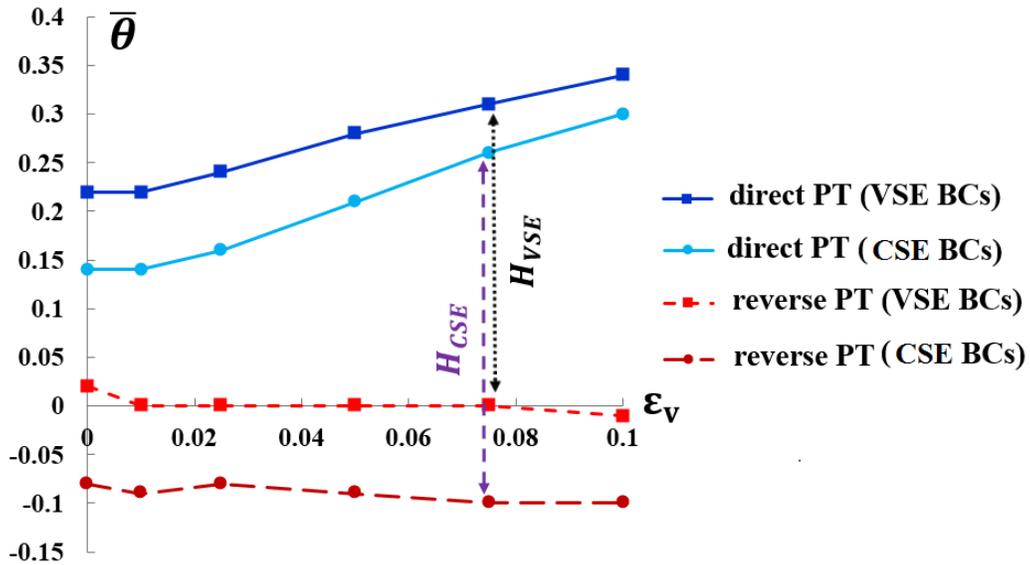

Fig. 5. The critical thermal driving forces vs. the misfit strain coefficient for the direct and reverse PTs for the CSE BCs and VSE BCs at the precipitate surface for R=2 nm and L=20. The hysteresis range is also shown.

The critical thermal driving force for both the direct and reverse PTs and for both the VSE and CSE BCs are plotted vs. the precipitate concentration in Fig. 6 for $\varepsilon_v = 0.1$. In contrast to the problem with the small precipitate radius of R=0.5, the critical thermal driving force nonlinearly increases vs. the precipitate concentration for the direct PT. For the reverse PT, however, the critical thermal driving force for the CSE BCs linearly increases vs. the precipitate concentration while it is almost independent of the precipitate concentration for the VSE BCs. Also, for any concentration, the VSE BCs result in higher critical thermal driving forces for both the direct and reverse PTs and a smaller hysteresis range compared to those of the CSE BCs.

From the computational point of view there are two choices to change the precipitate concentration: one is to keep the sample size while varying the precipitate radius and the other is to keep the precipitate radius constant and varying the sample size. The critical thermal driving forces for the two choices, i.e., (a) L=50 with varying precipitate radius from 1 to 6 and (b) R=2 with varying sample size from 15 to 100, are compared in Fig. 7 for $\varepsilon_v = 0.1$ and the CSE BCs. As can be seen, for smaller concentrations of $c \leq 0.01$, both choices coincide while for larger concentrations they differ by 10-15%. This difference means that the effect of the third scale parameter, the phase interface width, is significant.

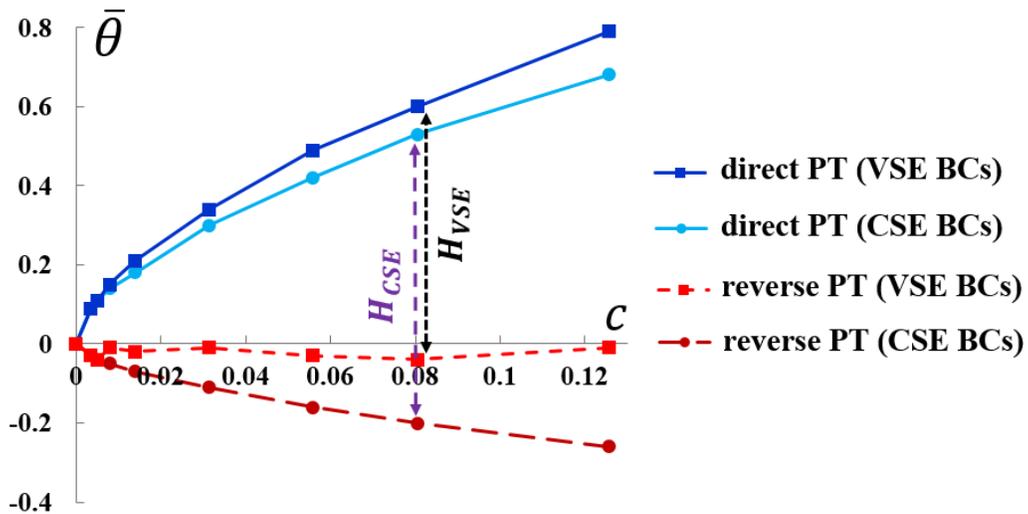

Fig. 6. The critical thermal driving forces for both the direct and reverse PTs and for both the VSE and CSE BCs vs. the precipitate concentration for $\varepsilon_v = 0.1$. R=2, 10<L<100

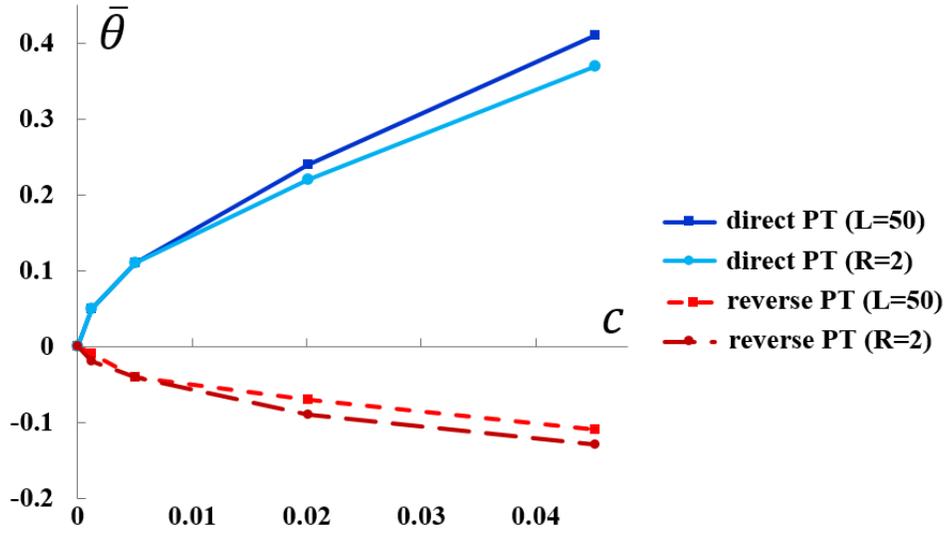

Fig. 7. The critical thermal driving forces for the direct and reverse PTs for two choices, i.e., (a) L=50 with varying precipitate radius and (b) R=2 with varying sample size, for $\varepsilon_v = 0.1$ and the CSE BCs.

Besides the variation of the critical athermal driving forces for the direct and reverse PTs with the concentration of the precipitate $c$ (Fig. 6), the morphology and the transformation rate also show a remarkable dependence on $c$. The dependence of the evolution of the HPP on $c$ is shown in Fig. 8 for the direct and reverse PTs for two different sample sizes of L=15 and 50 ($\varepsilon_v = 0.1$, R=2). Also, the phase concentration $\bar{\eta}$ vs. time is plotted for different sample sizes L=10, 12.5, 15, 20, 30, 40 and 50 for the direct PT in Fig. 9. For smaller sizes, i.e., larger precipitate concentrations, the effect of the precipitate is much larger so that the transformation rate during the interaction of precipitate and the interface is smaller (intermediate region). For larger sizes, this effect reduces so that such region disappears for L>30 and the concentration shows a linear variation, i.e., the transformation rate becomes the same before, during and after the interface passes the precipitate region. Obviously, for larger samples, the stationary solution is reached for larger times.

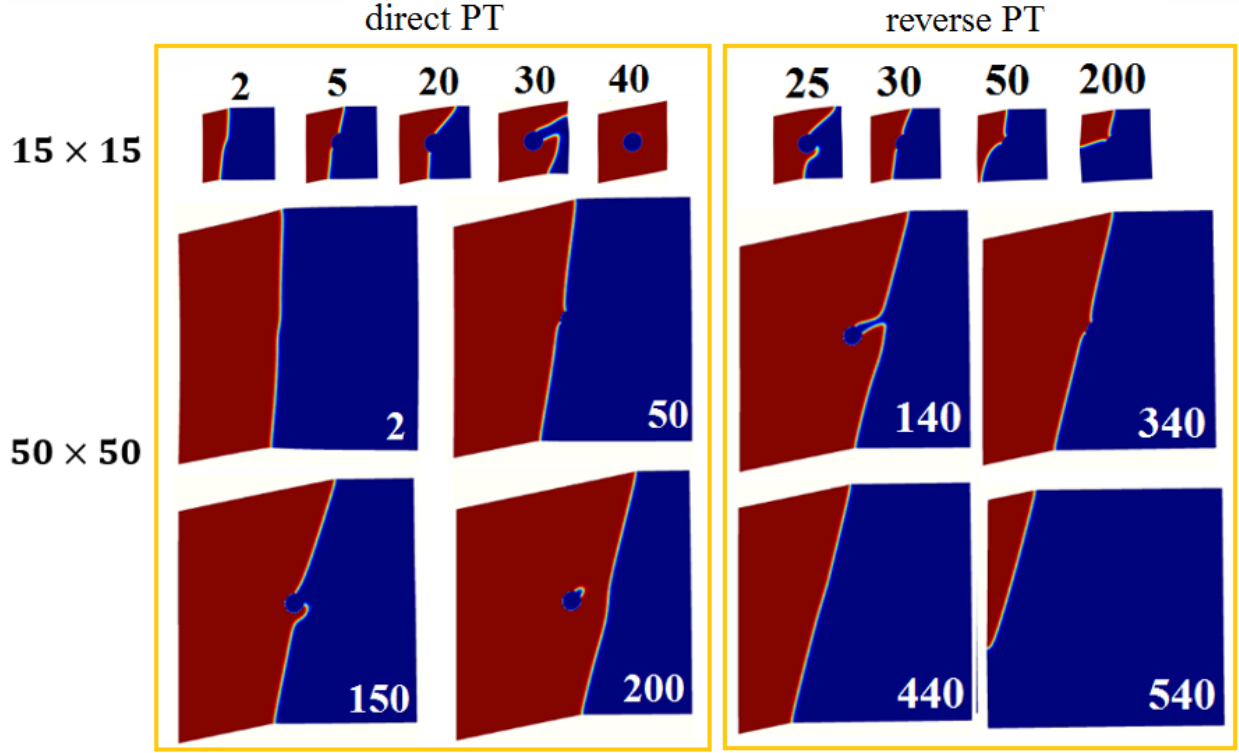

Fig. 8. The evolution of the HPP for the direct and reverse PTs for two different sample sizes of L=15 and 50 (R=2). For the direct PT, $\bar{\theta} = 0.42$ for L=15 and $\bar{\theta} = 0.12$ for L=50 (slightly larger than their critical thermal driving forces). For the reverse PT, $\bar{\theta} = -0.16$ for L=15 and $\bar{\theta} = -0.05$ for L=50 (slightly smaller than their critical thermal driving forces).

The pressure and transformation work ($W_{tr} = \boldsymbol{\sigma}:\boldsymbol{\varepsilon_{tr}}(\eta) - z(\theta - \theta_e)$) distributions corresponding to the thermal driving force $\bar{\theta} = 0.42$ for direct PT are plotted in Fig. 10 for both the CSE and VSE BCs for $c = 0.056$ and $\varepsilon_v = 0.1$. For a better illustration of the pressure inside the transforming region, the precipitate is excluded. Due to the VSE BCs, the precipitate is surrounded by the LPP and a continuous interface, and it is under pressure. For the CSE BCs, a part of the precipitate surface is surrounded by the HPP with a larger pressure concentration than for the VSE BCs; tensile stresses appear at the intersection of the interface and the precipitate and the boundary region between the precipitate surface and the LPP is under pressure.

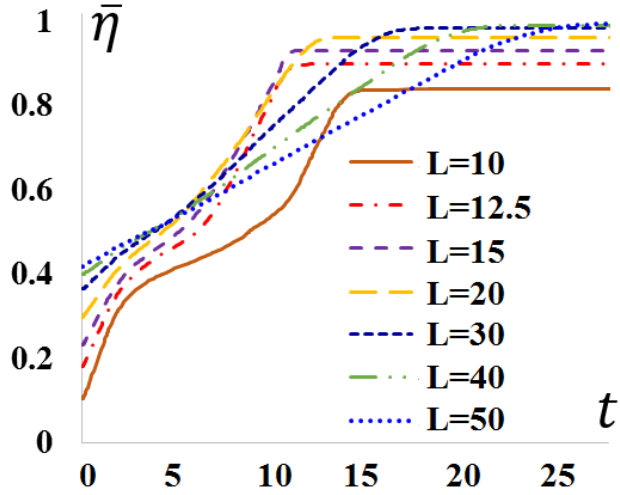

Fig. 9. The variation of the phase concentration $\bar{\eta}$ vs. for different sample sizes for the direct PT.

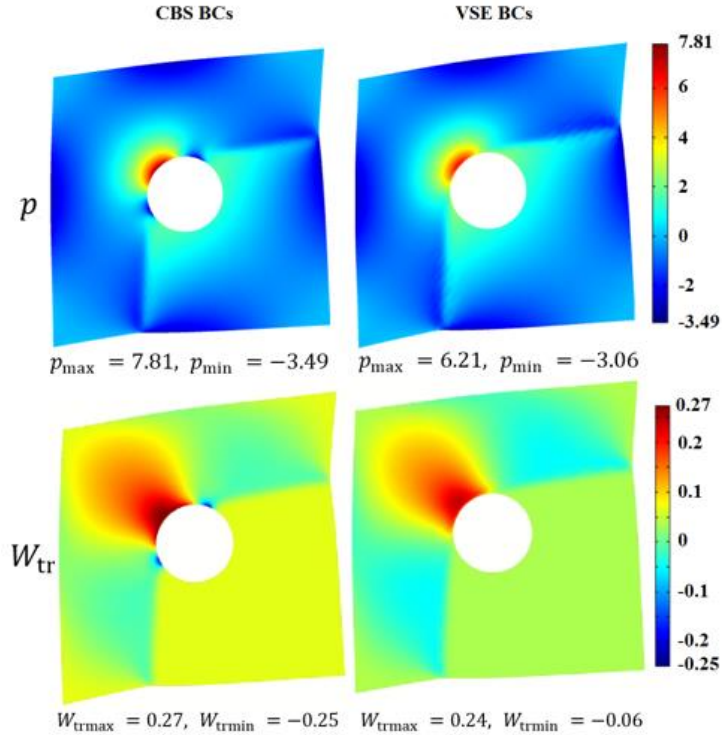

Fig. 10. The pressure distribution corresponding to the critical thermal driving force $\bar{\theta} = 0.42$ for direct PT for the CSE and VSE BCs for $c = 0.056$ and $\varepsilon_v = 0.1$.

The transformation work also shows higher concentrations around the precipitate region in the HPP for the CSE BCs ($\approx 0.25$) than for the VSE BCs ($\approx 0.2$) but almost the same low values along the interface away from the precipitate. The difference between the transformation work values for the two BCs is not significant; thus, a low difference is expected between the critical thermal driving forces for the CSE and VSE BCs ($\bar{\theta}_c^d = 0.42$ for the CSE BCs while $\bar{\theta}_c^d = 0.49$ for the VSE BCs).

The obtained critical thermal driving forces are validated using the local phase equilibrium condition criterion. To do so, the contour of the equilibrium condition $W_e = \boldsymbol{\sigma}:\boldsymbol{\varepsilon}_{tr} - z(\theta - \theta_e) = 0$ [18] is plotted on the stationary HPP nanostructure at the critical thermal driving force for both the direct and reverse PTs and the VSE and CSE BCs in Fig. 11. It is seen that the interface well coincides with the phase equilibrium condition contour $W_e = 0$, confirming thermodynamic phase equilibrium for the stationary solution.

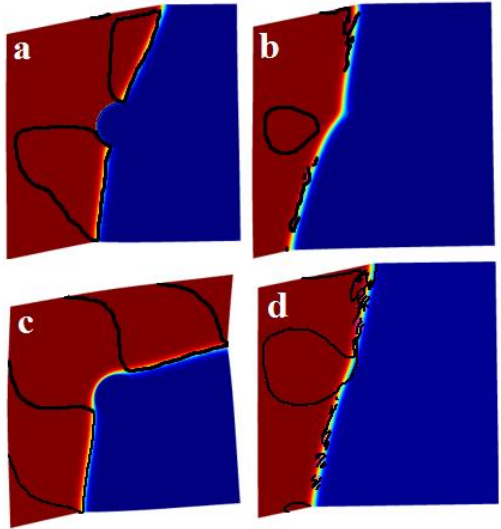

Fig. 11. The morphology of the HPP nanostructure at the critical thermal driving force for (a) the direct PT with the CSE BCs ($\bar{\theta}_c^d = 0.21$), (b) the reverse PT with the CSE BCs ($\bar{\theta}_c^d = -0.11$), (c) the direct PT with the VSE BCs ($\bar{\theta}_c^d = 0.28$), and (d) the reverse PT with the VSE BCs ($\bar{\theta}_c^r = -0.01$), combined with the contour line $W_e = 0$.

To better show the highly heterogeneous stress field and the transformation work, the evolution of the transformation work $W_{tr} = \boldsymbol{\sigma}:\boldsymbol{\varepsilon}_{tr}(\eta) - z(\theta - \theta_e)$ during the direct PT is presented in Fig. 12 for the CSE BCs at $\bar{\theta} = 0.22$ and for the VSE BCs at $\bar{\theta} = 0.29$. The high tensile and compressive stress concentrations and consequently, transformation work concentrations appear near the precipitate. The suppression of the interface motion at the precipitate occurs in the regions of negative transformation work (blue regions). Also, the interface motion occurs the regions where $W_{tr} > 0$. Since for each type of BCs the evolution is obtained for a slightly larger thermal driving force than its critical value, the transformation work along the moving interface regions is relatively low and mainly $0 < W_{tr} < 0.05$. As stated earlier, the VSE BCs result in the higher critical thermal driving force than the CSE BCs and consequently, the PT is promoted by the CSE BCs. This can be seen from the variation of the phase concentration with time for $\bar{\theta} = 0.3$ as an example. The phase concentration $\bar{\eta}$ is defined as the ratio of the transformed area to the total area. As can be seen in Fig. 13, the phase concentration for the CSE BCs is larger than that for the VSE BCs. The main difference between the two solutions appears during the time period $t_p$ when the interface is interacting with the precipitate and as it is clear, the rate of transformation is much larger for the CSE BCs. After this period, the transformation rate is again similar for both cases until they reach their stationary solutions. As a result, the stationary solution for the CSE BCs is reached at a shorter time ($t = 38$) than for the VSE BCs ($t = 55$).

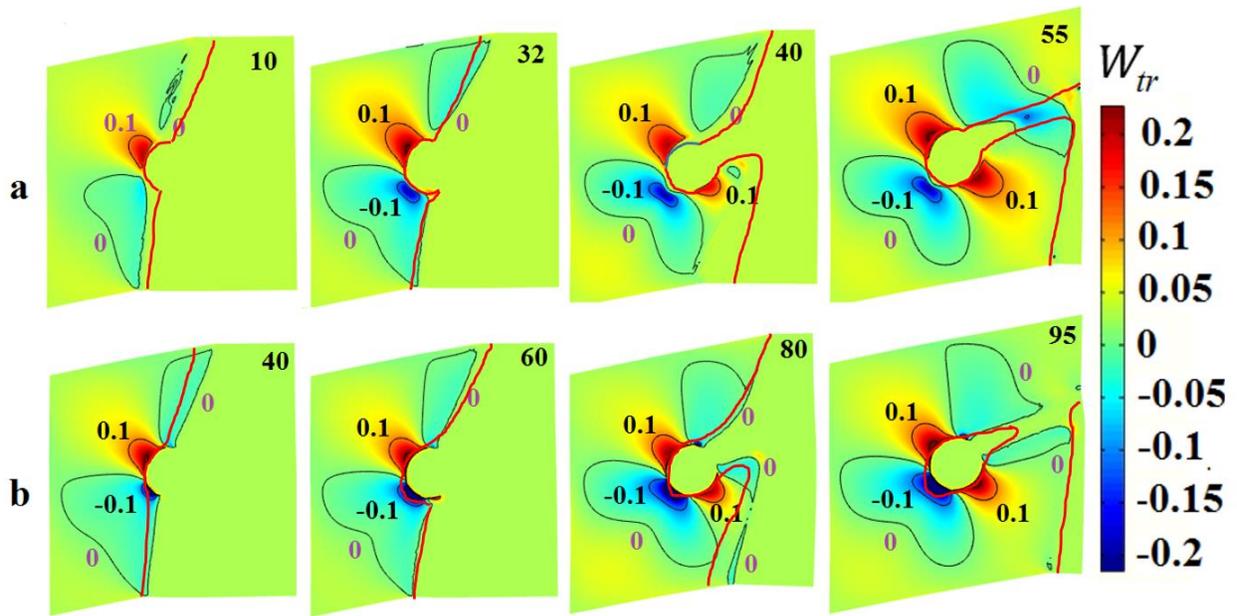

Fig. 12. The evolution of the transformation work $W_{tr}$ during the direct PT for the CSE BCs at $\bar{\theta} = 0.22$ (a) and for the VSE BCs at $\bar{\theta} = 0.29$ (b). The phase interfaces are included in each figure.

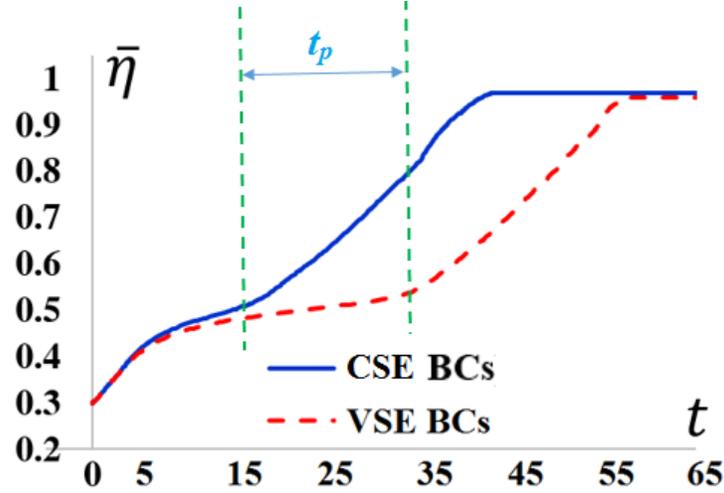

Fig. 13. The phase concentration $\bar{\eta}$ vs. time for the CSE and VSE BCs $\bar{\theta} = 0.3$.

It is worthy to note that the significant rotation of the stationary interface not only can be caused by the VSE BCs (like in Fig. 11), but also occurs for the large misfit strains even for the CSE BCs. Fig. 14 shows the distribution of $W_{tr}$ combined with the stationary interface contour lines of $\eta = 0.5$ for two different misfit strain coefficients $\varepsilon_v = 0.01$ and 0.1, for the direct PT for the CSE BCs, and a significant rotation of the interface is found for the large misfit strain $\varepsilon_v = 0.1$. The interface contour also corresponds to the regions of $W_{tr} = 0$ which proves its phase equilibrium state.

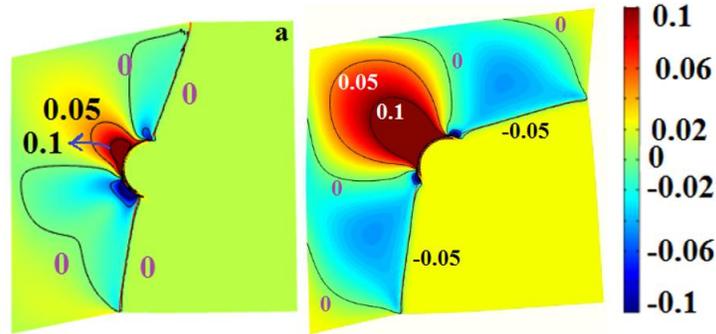

Fig. 14. The distribution of $W_{tr}$ combined with the stationary interface contour lines of $\eta = 0.5$ (red line) for $\varepsilon_v = 0.01$ at $\bar{\theta}_{th}^d = 0.14$ (a) and $\varepsilon_v = 0.1$ at $\bar{\theta}_{th}^d = 0.29$ (b) for the CSE BCs.

It is worthy to investigate the effect of interface width on the critical thermal driving force. In fact, by changing the interface width, the interaction of interface with precipitate surface changes which affects the critical thermal driving force. Here, the interface width is varied by changing parameter

$\beta$ and the critical thermal driving force is plotted vs. the normalized LPP-HPP interface width $\delta/\delta_0$ in Fig. 15 for direct PT for the CSE BCs at the precipitate surface and for $C = 0.056$ and $\varepsilon_v = 0.1$. As can be seen, the critical thermal driving force almost linearly reduces from 0.35 to 0.21 within the interface width range from $\delta/\delta_0 = 0.33$ (closer to the sharp interface approach) to 3. A similar dependence of the results on the interface energy is also found for other used misfit strains and precipitate concentrations.

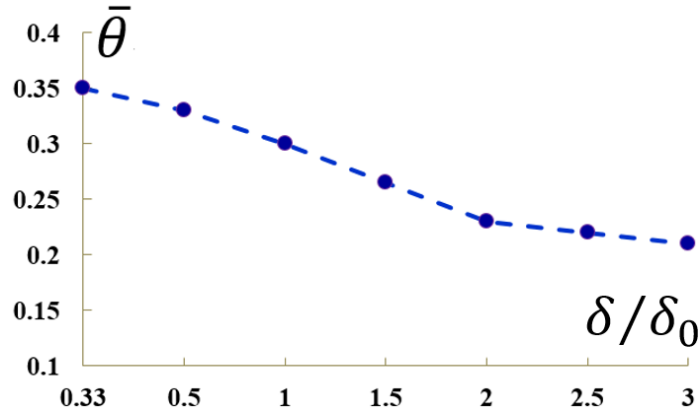

Fig. 15. The variation of the critical thermal driving force for direct PT (use subscripts th) vs. the normalized LPP-HPP interface width for the CSE at the precipitate surface and for $C = 0.056$ and $\varepsilon_v = 0.1$.

**Concluding remarks**

Athermal resistance to the LPP-HPP interface motion due to precipitates is investigated. The coupled phase field and elasticity equations are solved for the thermally-induced martensitic PT, which can also be interpreted as LPP-HPP PT, using the FEM code COMSOL. The volumetric misfit strain due to the precipitate is included using the error and constant functions. Due to the presence of a precipitate, the critical thermal driving forces remarkably differ between the direct and reverse PTs, resulting in a hysteresis behavior. The misfit strain does not practically show any effect on the critical thermal driving force if the precipitate radius is smaller than the interface width. The critical thermal driving forces nonlinearly increases vs. the precipitate concentration for both the direct and reverse PTs, especially for low precipitate concentrations. Change in the precipitate surface energy significantly changes the PT morphology and the critical thermal driving forces. In contrast to the CSE BCs, the interface does not split when passing the precipitate for the VSE BCs, but its width reduces. Also, the reverse PT reveals a different evolution compared to the direct PT, especially when passing the precipitate region. The critical thermal driving force shows dependence on the misfit strain for larger precipitate sizes than the interface width. For both the CSE and VSE BCs, the critical thermal driving force linearly increases vs. the misfit strain for the direct PT while it is almost independent of the misfit strain coefficient for the reverse PT. The VSE BCs also result in higher critical thermal driving forces for both the direct and reverse PTs and a smaller hysteresis range. In contrast to the problem with the small precipitate radii, the critical thermal driving force nonlinearly increases vs. the precipitate concentration for the direct PT. For the reverse PT, the critical thermal driving force for the CSE BCs linearly increases vs. the precipitate concentration, while it is almost independent of the precipitate concentration for the VSE BCs. Also, for any concentration, the VSE BCs result in higher critical thermal driving forces for both the direct and reverse PTs and a smaller hysteresis range. For a constant thermal driving force, the CSE BCs create a larger thermal driving force compared to the VSE BCs which results in a larger transformation rate during the interaction of the interface and the precipitate. The results of the two choices to change the precipitate concentration, i.e., the constant sample size with variable precipitate radius and the constant precipitate radius with variable sample size, coincide for smaller concentrations and slightly differ for larger concentrations. The obtained critical thermal driving forces are validated using the phase equilibrium condition criterion, so that the interfaces for the critical stationary morphologies well coincides with the zero thermodynamic driving force contour. The suppression of the interface motion at the precipitate is found at the regions of negative transformation work, while the interface motion belongs to the regions of positive transformation work. The morphology and the transformation rate also show a remarkable dependence on the sample size, so that for smaller sizes, the effect of the precipitate is much larger and the transformation rate during the interaction is smaller. The effect of the LPP-HPP interface width on the results was investigated. It is found that the critical thermal driving force weakly nonlinearly reduces as the LPP-HPP interface width increases. The obtained results help for a better understanding of athermal friction mechanism for interfaces and similar defect effects on various PTs at the nanoscale.

5. Acknowledgement

The support of Isfahan University of Technology and Iran National Science Foundation is for MJ gratefully acknowledged. VIL work was funded by NSF (MMN-1904830 and CMMI-1943710)

## 6. Declaration of interests

The authors declare that they have no known competing financial interests or personal relationships that could have appeared to influence the work reported in this paper.

**References**

[1] A. Cuniberti, S. Montecinos, F.C. Lovey, Effect of g2-phase precipitates on the martensitic transformation of a b-CuAlBe shape memory alloy, Intermetallics 17. (2009) 435–440.

[2] N. Zhou, C. Shen, M.F.-X. Wagner, G. Eggeler, M.J. Mills, Y. Wang, Effect of Ni4Ti3 precipitation on martensitic transformation in TiNi, Acta Mater. 58 (2010) 6685–6694.

[3] S. Y. Jiang, et. Al. Influence of $Ni_4Ti_3$ precipitates on phase transformation of NiTi shape memory alloy. T. Nonferr. Metal. Soc. 25 (12) 4063-4071. 2015.

[4] J. I. Pérez-Landazábal, V. Recarte, D. S. Agosta, V. Sánchez-Alarcos, and R. G. Leisure. Defect pinning of interface motion in thermoelastic structural transitions of Cu-Al-Ni shape-memory alloy. Phys. Rev. B 73, 224101.

[5] S. Ataollahi, M. J. Mahtabi, A molecular dynamics study on the effect of precipitate on the phase transformation in NiTi". ReSEARCH Dialogues Conference proceedings. https://scholar.utc.edu/research-dialogues/2021/posters/2.

[6] A. Basak, V. I. Levitas. Matrix-precipitate interface-induced martensitic transformation within nanoscale phase field approach: Effect of energy and dimensionless interface width. Acta Mater. 189, 2020 255-265.

[7] C. L. Gong, F. S. Han, Z. Li and M. P. Wang. Internal friction peak associated with the interface motion in the martensitic transformation of CuAlNiMnTi shape memory alloy. J. Appl. Phys. 102, 023521 (2007).

[8] G. Ghosh and G. B. Olson, Kinetics of FCC-BCC heterogeneous martensitic nucleation –I. The critical driving force for athermal nucleation. Acta. Mater. 42 (10) 3361-3370. 1994.

[9] V. I. Levitas, A.V. Idesman, G. B. Olson, and E. Stein, Numerical modelling of martensitic growth in an elastoplastic material. Philos. Mag. A 82, 429-462 (2002).

[10] V. I. Levitas, D. W. Lee. Athermal Resistance to Interface Motion in the Phase-Field Theory of Microstructure Evolution. Phys. Rev. Lett. 99. 245701.

[11] V.I. Levitas, D.-W. Lee, D.L. Preston, Interface propagation and microstructure evolution in phase field models of stress-induced martensitic phase transformations, Int. J. Plast. 26 (2010) 395–422.

[12] V. I. Levitas, M. Javanbakht. Phase field approach to interaction of phase transformation and dislocation evolution. Appl. Phys. Lett. 102, 251904 (2013).

[13] L. Wang, et al. Long-Range Ordered Carbon Clusters: A Crystalline Material with Amorphous Building Blocks. Science 337 (2012) 825-828.

[14] Y. Zhao, D.W. He, L.L. Daemen, et al. Superhard B–C–N materials synthesized in nanostructured bulks. J. Mater. Res. 17 (2002) 3139–3145.

[15] L. Q. Huston, A. Lugstein, J.S. Williams, J.E. Bradly, The high pressure phase transformation behavior of silicon nanowires. Appl. Phys. Lett. 113 (2018) 123103.

[16] V. I. Levitas, High pressure phase transformations revisited. J. Phys.: Condens. Matter.30 (2018) 163001.

[17] Levitas V.I. High-Pressure Phase Transformations under Severe Plastic Deformation by Torsion in Rotational Anvils. Material Transactions, 2019, Vol. 60, No. 7, 1294-1301.

[18] M. Javanbakht, V.I. Levitas, Nanoscale mechanisms for high-pressure mechanochemistry: a phase field study, J. Mater. Sci. 53 (2018) 13343–13363.

[19] M. Javanbakht, V.I. Levitas, Interaction between phase transformations and dislocations at the nanoscale. Part 2: Phase field simulation examples, J. Mech. Phys. Solids. 82 (2015) 164–185.

[20] M. Javanbakht, V.I. Levitas, Phase field simulations of plastic strain-induced phase transformations under high pressure and large shear, Phys. Rev. B. 94 (2016) 214104.

[21] M, Javanbakht, M.S, Ghaedi, Nanovoid induced multivariant martensitic growth under negative pressure: Effect of misfit strain and temperature on PT threshold stress and phase evolution. Mech Mater 2020:103627.

[22] V.I. Levitas, M. Javanbakht, Phase transformations in nanograin materials under high pressure and plastic shear: nanoscale mechanisms, Nanoscale. 6 (2014) 162–166.

[23] Couturier, G., et al. "Finite Element Simulations of 3D Zener Pinning." Materials Science Forum, vol. 467–470, Trans Tech Publications, Ltd., Oct. 2004, pp. 1009–1018.

[24] N. Fujiyama, et al. Austenite grain growth simulation considering the solute-drag effect and pinning effect. Sci Technol Adv Mater. 2017; 18(1): 88–95.

[25] Zhiqiang Li, Junsheng Wang, Houbing Huang, Influences of grain/particle interfacial energies on second-phase particle pinning grain coarsening of polycrystalline, J. Alloys Compd. 818(2020) 152848.

[26] R.K. Koju, K.A. Darling, K.N. Solanki, Y. Mishin, Atomistic modeling of capillary-driven grain boundary motion in Cu-Ta alloys, Acta Mater. 148(2018) 311-319.

[27] D. Chen, T. Ghoneim, Y. Kulkarni. Effect of pinning particles on grain boundary motion from interface random walk. Appl. Phys. Lett. 111, 161606 (2017).

[28] A.R. Eivani, et al. Effect of the Size Distribution of Nanoscale Dispersed Particles on the Zener Drag Pressure. Metall Mater Trans A 42 (2011) 1109–1116.

[29] M. Gandman, M. Ridgway, R. Gronsky, A. M. Glaeser, Microstructural evolution in Pt-implanted polycrystalline Al2O3, Acta Mater. 83 (2015) 169-179.

[30] H.L. Yang, et al., Study on recrystallization and correlated mechanical properties in Mo-modified Zr-Nb alloys, Mater. Sci. Eng. A, 661 (2016) 9-18.

[31] A. Graux, et al. Precipitation and grain growth modelling in Ti-Nb microalloyed steels,

Materialia, 5 (2019) 100233.

[32] Li Y, Zhou J, Li R and Zhang Q (2021) Molecular Dynamics Simulation of Zener Pinning by Differently Shaped and Oriented Particles. Front. Mater. 8: 682422.

[33] A.V. Mikhaylovskaya, M. Esmaeili Ghayoumabadi, A.G. Mochugovskiy, Superplasticity and mechanical properties of Al–Mg–Si alloy doped with eutectic-forming Ni and Fe, and dispersoid-forming Sc and Zr elements, Mater. Sci. Eng. A, 817 (2021).

[34] H. Dong, et al. Analysis of the interaction between moving α/γ interfaces and interphase precipitated carbides during cyclic phase transformations in a Nb-containing Fe-C-Mn alloy, Acta Mater. 158 (2018) 167-179.

[35] J. Wang, Effect of sintering temperature on phase transformation behavior and hardness of high-pressure high-temperature sintered 10 mol% Mg-PSZ. Ceramics Int. 47, 2021; 15180-15185.

[36] V.I. Levitas, H. Jafarzadeh, G.H. Farrahi, M. Javanbakht, Thermodynamically consistent and scale-dependent phase field approach for crack propagation allowing for surface stresses, Int. J. Plast. 111 (2018) 1–35.

[37] Y. Gao, et al. Formation and self-organization of void superlattices under irradiation: A phase field study. Materialia, 1 (2018) 78-88.

[38] Y. Li, D. Ma, B. Wang, Influence of bulk free energy density on single void evolution based on the phase-field method, Comput. Mater. Sci. 163 (2019) 100–107.

[39] Y. Wu, Q. Luo, E. Qin. Influencing factors of abnormal grain growth in Mg alloy by phase field method. Mater. Today. Commun. 22 (2020) 100790.

[40] M. Mamivand, M. Asle Zaeem, H. El Kadiri, L.Q. Chen. Phase field modeling of the tetragonal-to-monoclinic phase transformation in zirconia. Acta Mater. 61 (14) 5223-5235. 2013.

[41] A.E. Jacobs, S.H. Curnoe, R.C. Desai, Simulations of cubic-tetragonal ferroelastics, Phys. Rev. B. 68 (2003) 224104.

[43] D.J. Seol, S.Y. Hu, Y.L. Li, L.Q. Chen, K.H. Oh, Cubic to tetragonal martensitic transformation in a thin film elastically constrained by a substrate, Met. Mater. Int. 9 (2003) 221–226.

[44] S.Y. Hu, Y.L. Li, Y.X. Zheng, L.Q. Chen, Effect of solutes on dislocation motion —a phase-field simulation, Int. J. Plast. 20 (2004) 403–425..

[45] Y.U. Wang, Y.M. Jin, A.G. Khachaturyan, Phase field microelasticity modeling of dislocation dynamics near free surface and in heteroepitaxial thin films, Acta Mater. 51 (2003) 4209–4223.

[46] D. Rodney, Y. Le Bouar, A. Finel, Phase field methods and dislocations, Acta Mater. 51 (2003) 17–30.

[47] Y. Wang, J. Li, Phase field modeling of defects and deformation, Acta Mater. 58 (2010) 1212–1235.

[48] Schwarze, Christian, R. D. Kamachali and I. Steinbach. "Phase-field study of zener drag and pinning of cylindrical particles in polycrystalline materials." Acta Mater. 106 (2016): 59-65.

[49] K. Chang, J. Kwon, C.K. Rhee. Effect of particle-matrix coherency on Zener pinning: A phase-field approach. Comp. Mater. Sci. 142 (2018) 297-302.

[50] Tamoghna Chakrabarti, Sukriti Manna, Zener pinning through coherent precipitate: A phase-field study, Comp. Mater. Sci. 154(2018) 84-90.

[51] N. Moelans, B. Blanpain, P. Wollants, Phase field simulations of grain growth in two-dimensional systems containing finely dispersed second-phase particles, Acta Mater. 54 (4) (2006) 1175-1184.

[52] V. I. Levitas, D. L. Preston, Three-dimensional Landau theory for multivariant stress induced martensitic phase transformations. I. Austenite-martensite, Phys. Rev. B. 66 (2002) 134206.

[53] V. I. Levitas, D. L. Preston, Three-dimensional Landau theory for multivariant stress induced martensitic phase transformations. II. Multivariant phase transformations and stress space analysis, Phys. Rev. B. 66 (2002) 134207.

[54] V.I. Levitas, D.L. Preston, D.-W. Lee, Three-dimensional Landau theory for multivariant stress-induced martensitic phase transformations. III. Alternative potentials, critical nuclei, kink solutions, and dislocation theory, Phys. Rev. B. 68 (2003) 134201.

[55] V.I. Levitas, M. Javanbakht, Surface Tension and Energy in Multivariant Martensitic Transformations: Phase-Field Theory, Simulations, and Model of Coherent Interface, Phys. Rev. Lett. 105 (2010) 165701.

[56] M. Javanbakht, E. Barati, Martensitic phase transformations in shape memory alloy: phase field modeling with surface tension effect, Comput. Mater. Sci. 115 (2016) 137–144.

[57] S. Mirzakhani, M. Javanbakht, Phase field-elasticity analysis of austenite–martensite phase transformation at the nanoscale: Finite element modeling, Comput. Mater. Sci. 154 (2018) 41–52.

[58] Levitas V.I. and Javanbakht M. Phase-field approach to martensitic phase transformations: Effect of martensite-martensite interface energy. International Journal of Materials Research, 2011, Vol. 102, 652-665.

[59] H. Jafarzadeh, V.I. Levitas, GH. Farrahi, M. Javanbakht, Phase field approach for nanoscale interactions between crack propagation and phase transformation, Nanoscale 11 (46), 22243-22247.

[60] M. Mamivand, M.A. Zaeem, H. El Kadiri, A review on phase field modeling of martensitic phase transformation, Comput. Mater. Sci. 77 (2013) 304–311.

[61] Y. Wang, L. Peng, Y. Wu, Y. Zhao, Y. Wang, Y. Huang, W. Ding, Phase-field modeling the effect of misfit on the precipitation of the second-phase particles and grain coarsening, Comp. Mater. Sci. 100 (2015) 166–172.

[62] J. Kundin, H. Emmerich, J. Zimmer. Three-dimensional model of martensitic transformations with elasto-plastic effects, Philos. Mag. 90:11, 1495-1510, (2010).